\documentclass{pasj01}
\usepackage{natbib}

\usepackage{amsmath}
\Received{$\langle$11-Jan-2017$\rangle$}
\Accepted{$\langle$20-Sep-2017$\rangle$}
\Published{$\langle$publication date$\rangle$}

\begin{document}

\title{Temporal relations between magnetic bright points and the solar sunspot cycle}
\author{Dominik \textsc{Utz}\altaffilmark{1,2,3}, Richard \textsc{Muller}\altaffilmark{4} and Tom \textsc{Van Doorsselaere}\altaffilmark{2}}%
\altaffiltext{1}{IGAM/Institute of Physics, Karl-Franzens University Graz, Universit\"atsplatz 5/II, Graz 8010, Austria}
\altaffiltext{2}{Centre for mathematical Plasma Astrophysics, Mathematics Department, KU Leuven, Celestijnenlaan 200B, Leuven 3001, Belgium}
\altaffiltext{3}{Instituto de Astrof\'isica de Andaluc\'ia IAA-CSIC, Glorieta de la Astronom\'ia s/n, Granada 18008, Spain}
\altaffiltext{4}{Laboratoire d'Astrophysique de Toulouse et Tarbes, UMR 5572, CNRS et Universit\'e Paul Sabatier Toulouse 3, 57 avenue d'Azereix, Tarbes 65000, France}

\email{Dominik.Utz@uni-graz.at}


\maketitle

\begin{abstract}
The Sun shows a global magnetic field cycle traditionally best visible in the photosphere as a changing sunspot cycle featuring roughly an 11 year period. In addition we know that our host star also harbours small-scale magnetic fields often seen as strong concentrations of magnetic flux reaching kG field strengths. These features are situated in inter-granular lanes where they show up bright as so-called magnetic bright points (MBPs). In this short paper we wish to analyse a homogenous nearly 10 year long synoptic Hinode image data set recorded from November 2006 up to February 2016 in the G-band to inspect the relationship between the number of MBPs at the solar disc centre and the relative sunspot number.

Our findings suggest that indeed the number of MBPs at the solar disc centre is correlated to the relative sunspot number, but with the particular feature of showing two different temporal shifts between the decreasing phase of cycle 23 including the minimum and the increasing phase of cycle 24 including the maximum. While the former is shifted by about 22 months the later is only shifted by less than 12 months. Moreover, we introduce and discuss an analytical model to predict the number of MBPs at the solar disc centre purely depending on the evolution of the relative sunspot number as well as the temporal change of the relative sunspot number and two background parameters describing a possibly acting surface dynamo as well as the strength of the magnetic field diffusion. Finally, we are able to confirm the plausibility of the temporal shifts by a simplistic random walk model.

The main conclusion to be drawn from this work is that the injection of magnetic flux, coming from active regions as represented by sunspots, happens on faster time scales than the removal of small-scale magnetic flux elements later on.
  
\end{abstract}

\section{Introduction}
An important feature of our host star - the Sun - is its magnetic cycle visible as changing activity cycle throughout the whole solar atmosphere \citep[see, e.g.,][]{2010LRSP....7....1H}. While the varying global magnetic fields cause a lot of turbulent energy releases in the higher solar atmosphere like flares and CMES \citep[e.g.,][]{1994JGR....99.4201W} the most evident feature of the solar magnetic field cycle are the changing sunspot patterns. The variation of sunspots in number as well as in configuration must have been known for quite a long time but it was \citet[][]{1844AN.....21..233S} who first realised that the variation in number follows certain patterns and especially a period of about 11 years. \citet[][]{1908ApJ....28..315H} realized that sunspots are magnetic and thus it was established that there is a global magnetic field dynamo acting on the Sun \citep[e.g.,][]{1961ApJ...133..572B}. Recent progress in dynamo theories are now often driven by helioseismology on the on hand \citep[see, e.g.,][]{2014SSRv..186..191B}, which is practically the only method to look into the subsurface configuration of the Sun, and on the other hand by simulations \citep[e.g., a recent review by][]{2014SSRv..186..561K}.

In addition to this large scale dynamics of strong and extended magnetic fields we know of the existence of plenty of small-scale magnetic features \citep[see][]{1987ARA&A..25...83Z}. Among them are so-called magnetic bright points \citep[MBPs; e.g.,][]{2001ApJ...553..449B,2010ApJ...725L.101A}. Most often they are investigated by filtergram observations in the G-band \citep[see][]{2003ApJ...597L.173S,2013Ap&SS.348...17F}, but also covered by spectropolarimetric means \citep[e.g.,][]{2012SoPh..280..407R}, as well as recently by numerical approaches \citep[among others][]{2014A&A...568A..13R}. The interest in them arises due to their importance as they do not only represent a significant part of the small-scale magnetic flux budget \citep[e.g.,][]{1997ApJ...487..424S,2016ApJ...820...35G} due to their large magnetic field strength in the range of a kG \citep{2013A&A...554A..65U}, but additionally they contribute to the coronal heating problem via the creation and propagation of MHD waves into the higher solar atmosphere \citep[e.g.,][]{1993ApJ...413..811C,1994A&A...283..232M,2009Sci...323.1582J,2013SSRv..175....1M}.

In this particular contribution we would like to investigate the link between the large-scale magnetic cycle visible via sunspots and the small-scale magnetic field cycle via investigating the number of MBPs at the solar disc centre. While there has been plenty of information gathered about the sunspot cycle over the years, much less is known about the variation of small-scale magnetic fields. This is especially true for MBPs \citep[see e.g.,][]{1984SoPh...94...33M}. A good overview of earlier attempts of establishing relations between small-scale solar flux elements and the sunspot cycle is given in the work of \citet{2011ApJ...731...37J} within the opening chapter. In an earlier paper of the current authors \citep{2016A&A...585A..39U} we found an in phase correlation between the number of MBPs and the sunspot cycle, which was shifted by about 2.5 years with the sunspots being ahead of the corresponding MBP number. In the current work we wish to re-investigate the same, but updated, data set (described in Section 2) which is now more complete and extended by several months and thus also covers the solar maximum of the current cycle 24 (full analysis see Section 3), while the older work was only relying on the minimum of cycle 23 and the early rise phase of cycle 24. Moreover, we also wish to investigate in much more detail the contribution of the sunspots to the number of MBPs detected at the disc centre and model this influence (Section 4). In the following Section 5 we show the usefulness of the model for predictions of the MBP activity close to the equator before we finish this small paper with a discussion of what can be learned from the modeling efforts in Section 6 and then give our conclusions and outlook in the final Section 7.

\begin{figure*}
 \begin{center}
  \includegraphics[width=0.85\textwidth]{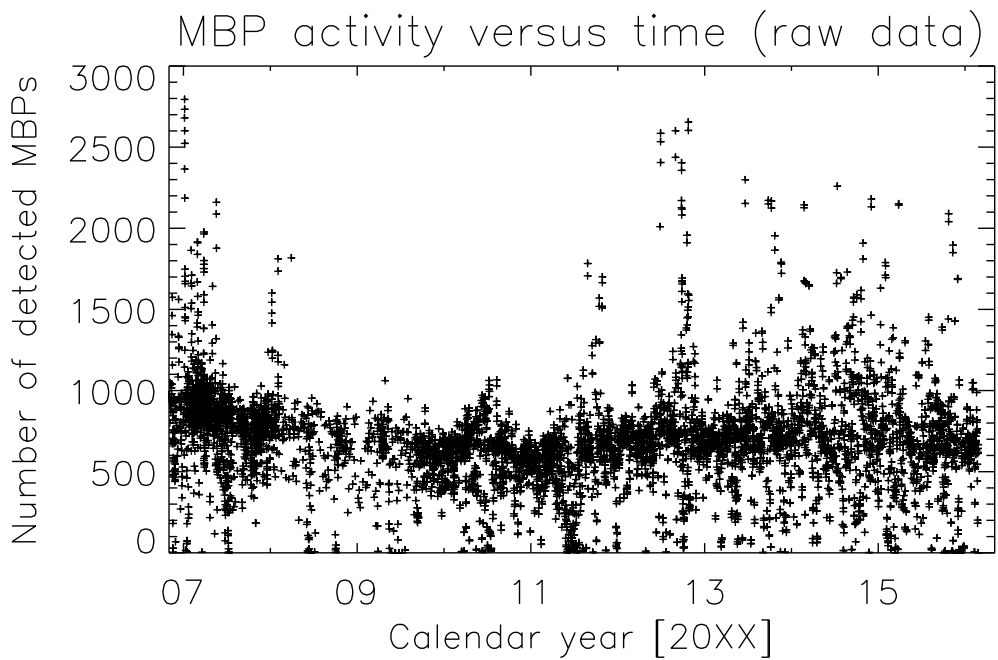}
 \end{center}
 \caption{raw results; The number of detected MBPs in all the analysed Hinode G-band images is shown versus time. This is an updated and extended representation of Fig. 2 of the work \citet{2016A&A...585A..39U}.}\label{fig1}
\end{figure*}

\section{Data and basic analysis}
In this study we used the Hinode/SOT synoptic filtergram data taken at the solar disc centre. In the early phase of the spacecraft mission the data was obtained on a daily basis while the amount of data taken was reduced after problems occurred with the downlink antenna. Within this programme images are taken in principle with all 6 broadband filters of the Hinode/SOT/BFI \citep[Solar Optical Telescope; Broadband Filter Imager;][]{2007SoPh..243....3K,2008SoPh..249..167T} instrument, which observes the blue, green, and red continuum as well as the CaII H line by a corresponding filter. Two more filters are available for molecular lines, namely the G-band centred around 4305 \AA, and the Cn band-head centred around 3883.5 \AA. In the beginning the data were taken with the highest possible sampling resolution of 0.054 arcsec/pixel with a full field of view of 4048 by 2024 pixels corresponding to about 220 by 110 arcsec$^2$. After the failure of the main downlink antenna the operators decided to perform an onboard binning of the data to reduce the amount of data to be downloaded to the ground control station. Thus the data taken after 17 February 2008 were already reduced in pixel sampling on-board to a value of 0.108 arcsec/pixel. To have a homogenous data set we decided to implement also a down-sampling of the earlier data recorded already previously to this date.

For the analysis and investigation of MBPs a preferential spectral region is the so-called G-band which covers molecular lines. These molecular absorption lines happen to be weakened in small-scale strong magnetic field agglomerations and thus such field concentrations can be observed with a higher contrast compared to the normal continuum \citep[see, e.g.,][]{2003ApJ...597L.173S}. Therefore, we selected all available 6343 G-band images covering the period from November 2006 up to February 2016. Unfortunately, a problem occurred in the mid of February 2016 which lead to a malfunctioning CCD and finally to the loss of the BFI and NFI instrument onboard of the Hinode spacecraft as both of the instruments use the same two CCD chips. Thus the synoptic programme cannot be prolonged anymore and we have therefore the longest available and most stable data set ever recorded from space for such a purpose at hand.

The principle data calibration follows now the idea already outlined in \citet[][]{2016A&A...585A..39U}. First we calibrate the data (flat fielding, dark current calibration) with the SSWIDL routine \verb+fg_prep+. In the next step we apply an automated image segmentation and MBP identification algorithm as described in detail in \citet[][]{2009A&A...498..289U,2010A&A...511A..39U}. After applying the algorithm to all 6343 images we obtain the following graph as shown in Fig. \ref{fig1}.

\begin{figure*}
 \begin{center}
  \includegraphics[width=0.85\textwidth]{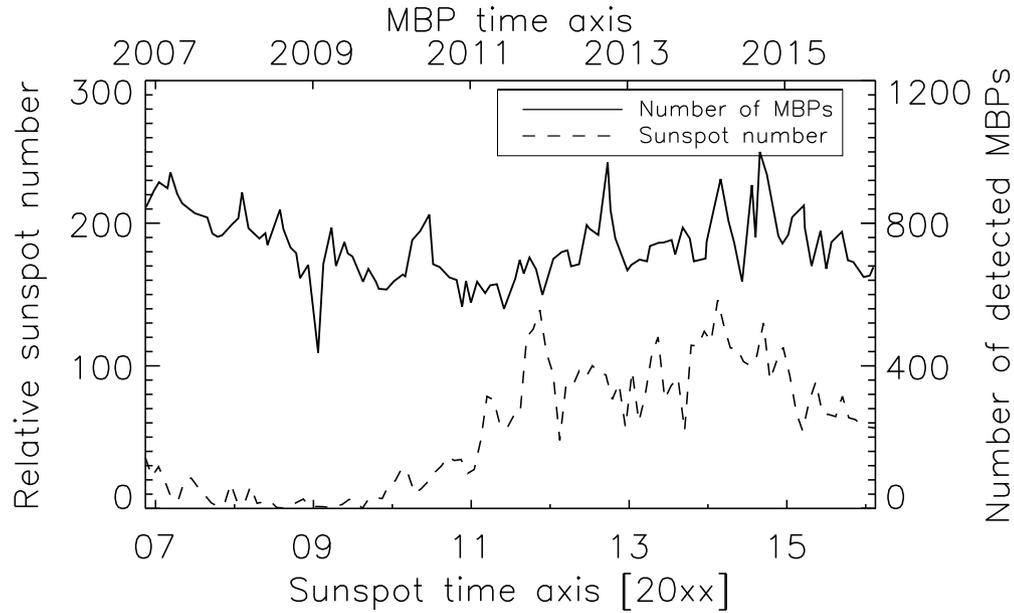}
 \end{center}
 \caption{The evolution of the median MBP number per month is plotted together with the evolution of the relative sunspot number for the decreasing solar cycle 23 and the increasing phase of the solar cycle 24 including both the minimum between the cycles and the maximum of the new ongoing cycle. This is an updated plot of Fig. 3 upper panel of the work \citet{2016A&A...585A..39U}.}\label{fig2}
\end{figure*}

On average the number of detected MBPs in the disc centre of the Sun is around 1000 corresponding to roughly 4 per 100 arcsec$^2$ (which would correspond to typical mesogranular diameters). Besides, it is visible that there are strong peaks reaching up to values of 2 to 3 thousand, i.e., more than doubling the usually visible bright point activity. These outliers are related to images in which sunspots are visible. These sunspots lead to an enhanced magnetic network and thus to a higher detection rate of MBPs. Furthermore, there are measurement points clearly below the 800 to 1000 level. These points are related to images out of focus and/or broken images, e.g., only a part of the image was successfully downloaded from the satellite\footnote{For a detailed analysis and discussion of the data stability (which was excellent) we wish to refer at this point the interested reader to \citet{2016A&A...585A..39U}.}.
To dispose most of these active region and broken images and thus to obtain a clearer insight into the overall behaviour we applied a two step selection and smoothing criterion. In the first step we applied a contrast criterion and selected only those images which have a contrast close to the optimum focal position which we defined as 10.8 to 11.8 \% (standard deviation of the image intensity divided by the mean image intensity times 100). After selecting thus only the best focused images, which already helps to dispose sunspot images and images out of focus, we calculated a median MBP number per month for the selected images of each monthly period. The result is depicted in Fig. \ref{fig2}. For completeness we wish to mention that on average 73\% of the images taken in a month survive our stringent selection method and that on average the median MBP number is then calculated over 41 images (for most months the number ranges between 25 to 60). In the least statistical robust month still at least 4 images were considered for the calculation of the median monthly MBP number. This low confidence month occurred in the period spanning the time from roughly 2008 to 2010, when Hinode was not taking as many images within the synoptic programme as usual (see also Fig. \ref{fig1}).

\section{Results}
Figure \ref{fig2} illustrates in solid line the number of MBPs detected in the Hinode synoptic G-band images at the disc centre after calculating a corresponding median number for each month. In dashed line we show the corresponding relative sunspot number obtained from the SIDC (Solar Influence Data Centre: \verb+http://sidc.oma.be/+). A similar result, but for a limited data set (data only available up to August 2014), was shown already in \citet{2016A&A...585A..39U}. In this earlier paper we found that the MBP cycle is shifted by about 2.5 years from the sunspot cycle. We wish to repeat this analysis with the extended data set now at hand which also covers the maximum of the cycle 24 which was missing in the earlier work.

\begin{figure}
 \begin{center}
  \includegraphics[width=0.9\columnwidth]{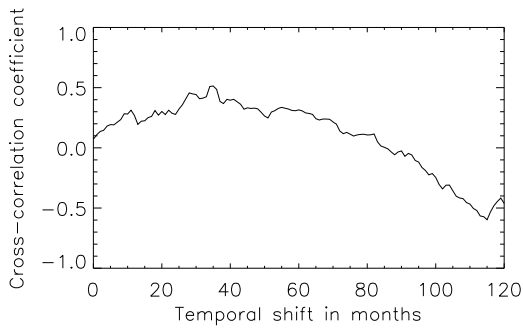}
 \end{center}
 \caption{the cross-correlation between the sunspot cycle as given by the relative sunspot number and the number of MBPs measured at the disc centre. The indicated monthly shifts correspond to the number of months the sunspot cycle would be ahead of the MBP cycle for the calculated and corresponding correlations.}\label{fig3}
\end{figure}

Thus we wish to have a detailed look on the correlation between the relative sunspot number and the detected monthly median number of MBPs. The resulting plot for the cross-correlation coefficient versus temporal shift between both quantities is depicted in Fig. \ref{fig3}.

Here we can identify more or less 3 local maxima, with the largest being situated at a temporal shift of 35 months between the sunspots and the number of MBPs. This would mean that the sunspots cycle is nearly 3 years ahead of the MBP activity at disc centre. The other two maxima would correspond to 28 months (the previously found 2.5 years lag) and only 11 months.

\begin{figure}
 \begin{center}
  \includegraphics[width=0.9\columnwidth]{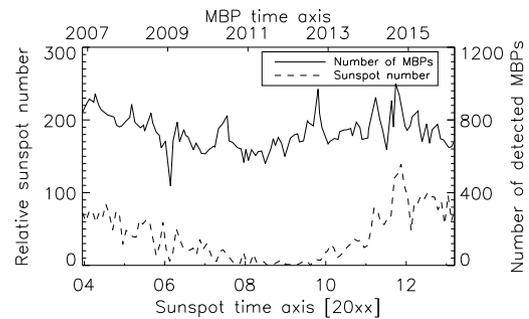}
 \end{center}
 \caption{gives the number of detected MBPs at the disc centre (solid line) together with the evolution of the relative number of sunspots. This plot shows the sunspot axis shifted by 35 months (dashed line) which corresponds to the general maximum of the cross-correlation (0.51) between both time series.}\label{fig4}
\end{figure}

Plotting now the sunspots shifted for the nearly 3 years period (35 months) yields Fig. \ref{fig4}. It becomes clear that the 3 year period is most likely an artifact caused by the one large variation around the solar cycle maximum at the end of the observational period. On the other hand it becomes clear that while their is an acceptable good agreement between the rising phase and maximum of sunspot cycle 24 with the rising activity of MBPs at the disc centre, the decreasing phase of the previous cycle, and especially the minimum between both cycles, shows a clear mismatch when the shifting period would be around 3 years. Thus we wish now to investigate the shifts for the decreasing cycle 23 and the rising phase of cycle 24 separately.  

\begin{figure}
 \begin{center}
  \includegraphics[width=0.9\columnwidth]{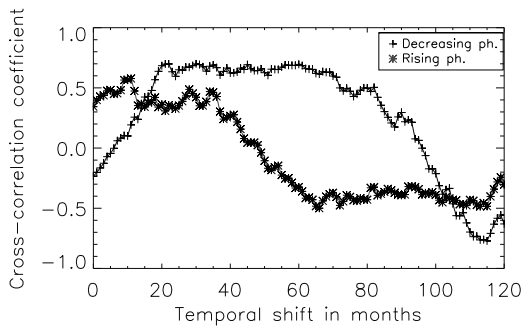}
 \end{center}
 \caption{cross-correlations between the number of MBPs during the rising phase of the sunspot cycle with the rising phase of sunspot cycle 24 (stars) and number of MBPs during the decreasing phase of cycle 23 with the corresponding part of the relative sunspot number during the decreasing phase (crosses). The selected periods as well as the time series plots with temporal shifts featuring the maximum correlations are shown in Fig. \ref{fig6} (for the rising phase) and Fig. \ref{fig7} (decreasing phase), respectively.}\label{fig5}
\end{figure}

This approach yields Fig. \ref{fig5}. Immediately it becomes clear that the rising phase of the sunspot cycle (correlation shown by stars) is nearly co-aligned with an increase in MBP activity. There is at most a shift of a few months, e.g. the correlation coefficient starts with 0.34, reaches a first local maxima of 0.48 after 6 months and the absolute maximum of the correlation is reached after 11 months, or roughly a year, with a value of 0.58. Thus the injection of new magnetic field into the network region follows strongly and temporally very closely within a year the sunspot cycle. On the other hand the picture is quite different for the decreasing phase of solar cycle 23. The correlation for this part of the period is shown by crosses and one can see that the time lag is much more pronounced. Here we reach the maximum correlation after a shift of 22 months with a value of 0.7, which means that the network takes quite some time to dispose the magnetic field injected from the previous sunspot cycle.

\begin{figure}
 \begin{center}
  \includegraphics[width=0.9\columnwidth]{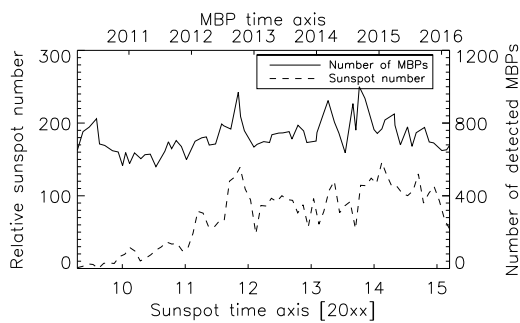}
 \end{center}
 \caption{the number of MBPs at disc centre (solid line) together with the relative sunspot number (dashed line) during the rising phase of cycle 24. The shift between both curves is 12 months and the correlation coefficient amounts to 0.58.}\label{fig6}
\end{figure}

\begin{figure}
 \begin{center}
  \includegraphics[width=0.9\columnwidth]{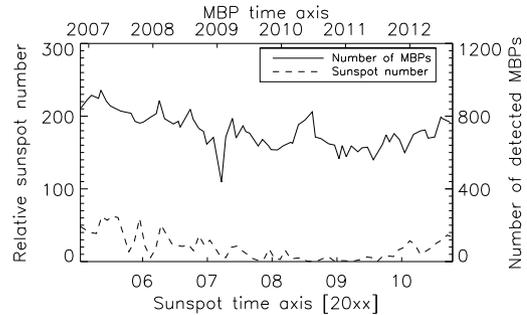}
 \end{center}
 \caption{similar to Fig. \ref{fig6}, but for the decreasing phase of cycle 23. The shift between both plotted curves is 22 months and the correlation coefficient reaches a maximum value of 0.7.}\label{fig7}
\end{figure}

The next two Figs. \ref{fig6} and \ref{fig7} depict the used periods for the rising and decreasing phase of the solar cycle with the activity of MBPs at disc centre shifted accordingly to the previously found best correlation values.

\begin{figure}
 \begin{center}
  \includegraphics[width=0.9\columnwidth]{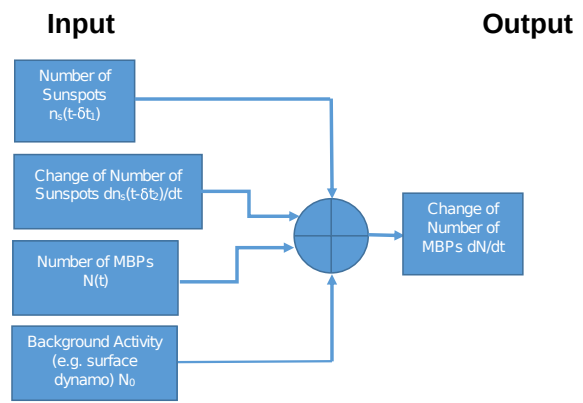}
 \end{center}
 \caption{electronic circuit schematic representing the governing differential equation which controls the change of the number of MBPs at the disc centre.}\label{fig8}
\end{figure}

\section{Modeling of the temporal behaviour of MBPs}
In the previous section we have seen that the injection of new flux into the network due to the rising phase of the new solar cycle 24 occurs on a much faster time scale with a shift of only a few months up to a year while the disposal of the flux within the magnetic network takes a longer time of about 2 years (22 months). Thus one may wonder if it is possible to construct a model to predict the change of magnetic network activity as seen by MBPs in the solar disc centre due to the sunspot activity as measured by the relative sunspot number.

After some reflections we would propose the following governing equation for the change of the number of MBPs $N(t)$ with time at the solar disc centre:
\begin{equation}
\frac{dN(t)}{dt}\propto a n_s(t-\delta t_1) +b \frac{dn_s(t-\delta t_2)}{dt} + c (N(t) - N_0)  
\label{principle}
\end{equation}
{The idea behind this conceptual equation is that the change in the number of MBPs will be proportional to the number of sunspots $n_s$ expressed by the proportionality parameter $a$, to the monthly derivative of the number of sunspots (the rate of sunspot change) $dn_s/dt$ expressed by parameter $b$\footnote{This parameter is justified by the idea that when a sunspot emerges there will be an immediate transport of new magnetic flux to the network.}, naturally to the amount of available MBPs related to the coefficient $c$ (which can be interpreted similar to a radioactive decay constant\footnote{The ``decay'' - flux dispersion - will depend (be proportional) on the amount of magnetic flux in place and can be caused, e.g., by the meridional flows transporting flux away from the disc centre to the poles.}). Having in mind that there might be a background solar surface dynamo working on, e.g., the granulation scale, we should reduce the number of effected MBPs in the equation above by a background permanent ``rest-flux'' number $N_0$\footnote{While the meridional flows will sweep magnetic field away from the disc centre (constant $c$), the background surface dynamo will ``replenish'' the network with new magnetic fields leading ultimately to some balanced number of MBPs -- the constant $N_0$.}}. This equation can be graphically illustrated as a flowchart such as Fig. \ref{fig8}. Another aspect which should be mentioned here is that the influence of the sunspot number as well as its change might be shifted in time by a certain amount, which is expressed as $\delta t_1$ and $\delta t_2$ in the equation above. This is a reasonable assumption as we have learned in the previous section that the number of detected MBPs generally lags behind the number of sunspots as well as that the decay of flux seems to happen slower than the injection of new flux. 

\begin{figure*}
 \begin{center}
  \includegraphics[width=0.9\textwidth]{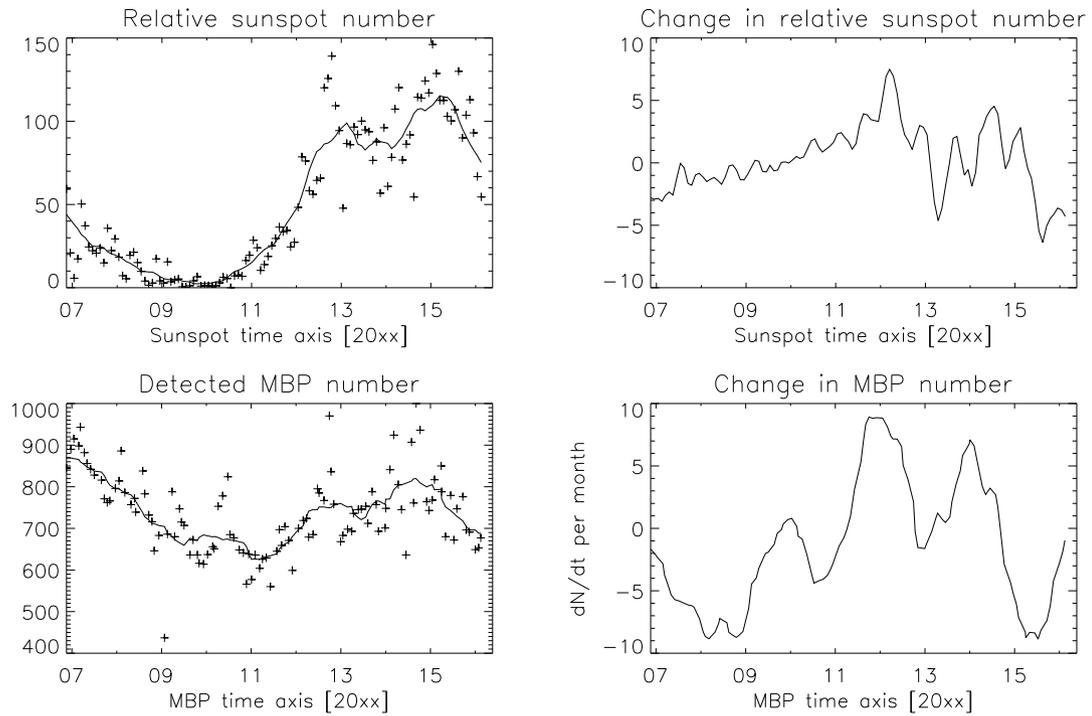}
 \end{center}
 \caption{the principle input components/curves for the differential equation (see, Eq. \ref{principle}) governing the change of the MBPs at disc centre (see also schematic Fig. \ref{fig8}). From left top to bottom right: relative sunspot number (solid line depicts a 12 month running average), the monthly derivative of the relative sunspot number calculated from the 12 month running averaged relative sunspot number as shown before, the number of detected MBPs at disc centre (solid line depicts again a 12 month running average), and finally the to be modeled output curve, the change of MBPs at disc centre calculated via averaging the number of MBPs over a yearly mean and then calculating the derivative of this curve and applying another running 12 month average. Originally measured data points are shown as crosses.}\label{fig9}
\end{figure*}

Thus our approach is based on the idea of constructing the profile and evolution of the $dN/dt$ temporal evolution as depicted in Fig. \ref{fig9} lower right bottom by three input variables, the sunspot number, the monthly derivative of the sunspot number (sunspot change rate), and the number of bright points itself. The temporal evolutions of these input variables are shown from the left top, relative sunspot number, via the right top, the monthly derivative of the relative sunspot number, to the left bottom, the actual detected number of MBPs. The four given quantities are depicted as running yearly average curves (solid line) and for the number of sunspots and the number of MBPs also the actual measured values are depicted by crosses.



\begin{figure*}
 \begin{center}
  \includegraphics[width=0.9\textwidth]{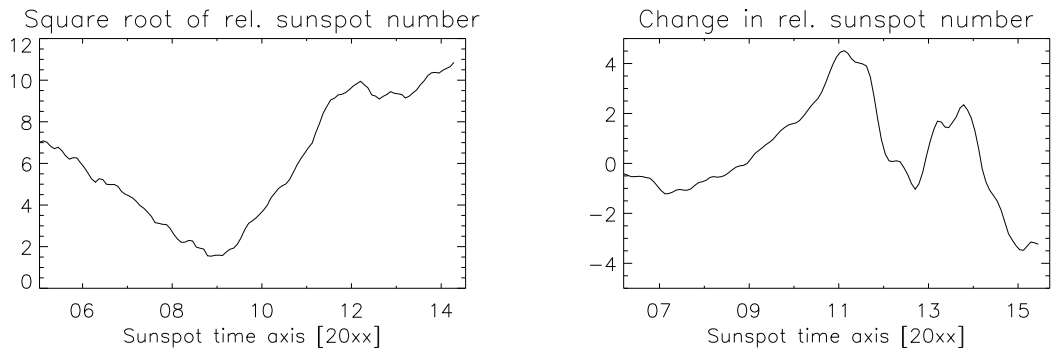}
 \end{center}
 \caption{the actually obtained best fitting candidates for the input parameters of the predictive model. From left top to right bottom: the square root of the relative sunspot number for fitting the change in the number of MBPs (shifted by 22 months) and the monthly derivative of the square root of the relative sunspot number (shifted by 8 months). Both curves are shown as yearly running smooths and thus slightly different from the ones depicted in the Fig. \ref{fig9}. For the obtained fitting parameters (weights of the different curves) see Table \ref{tab1}.}\label{fig11}
\end{figure*}

\begin{figure}
 \begin{center}
  \includegraphics[width=0.9\columnwidth]{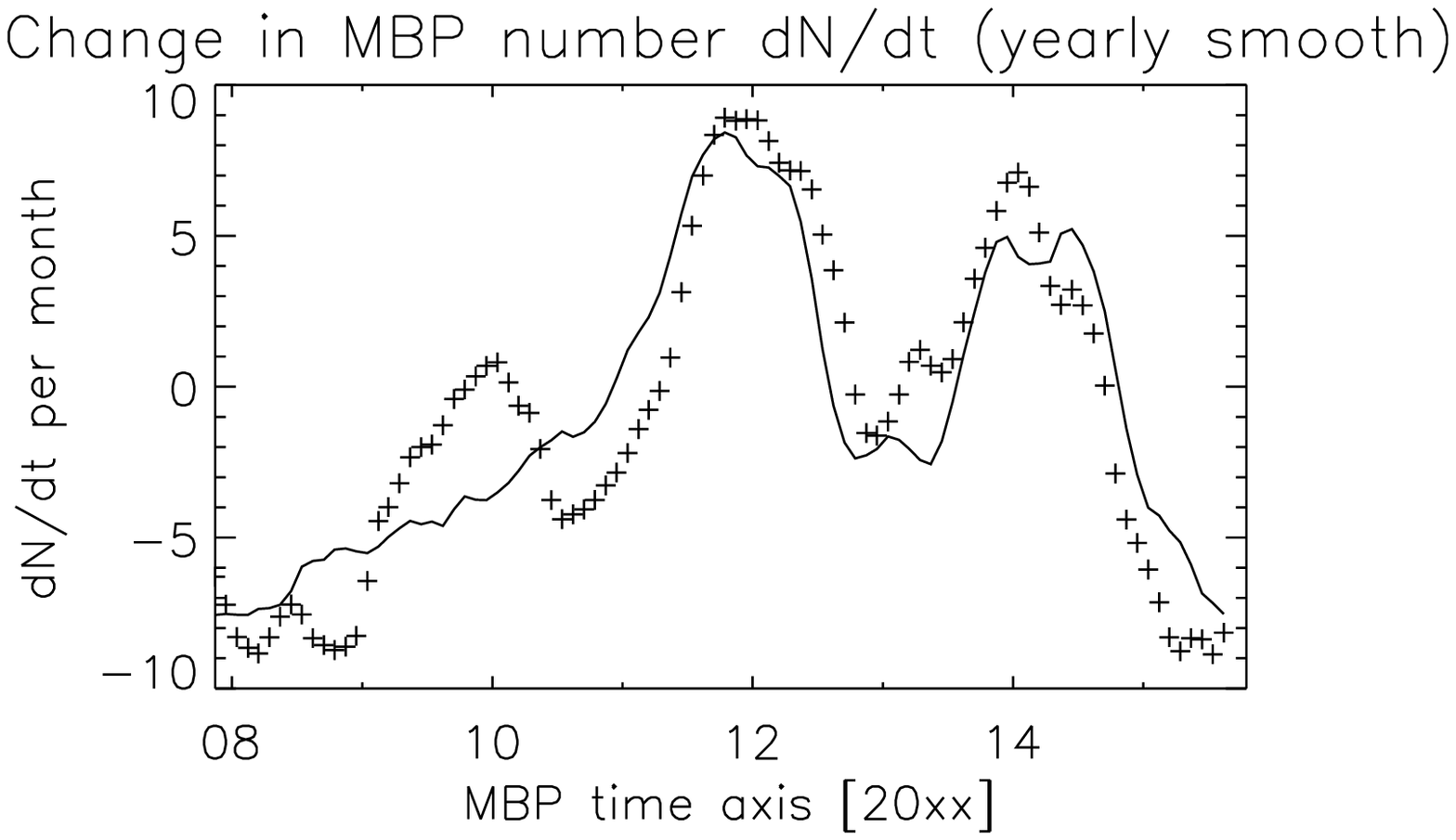}
 \end{center}
 \caption{the detected change of the number of MBPs after applying an annual running mean filtering shown with crosses and the fitted change of MBPs according to Eq. \ref{compl}. for the longest possible data set period from May 2007 to August 2015.}\label{fig12}
\end{figure}

After thorough testing of our approach we learned that actually an improved fit can be achieved when one uses rather the square root of the number of sunspots as input parameter then the sunspot number itself. This might be understood in the way that already a small number of sunspots can contribute significantly to the number of MBPs at the disc centre, while a higher number of sunspots start to saturate the magnetic network, or, at least, lead to a decreasing impact on the number of detected MBPs. Thus, the finally applied equation has been (already taking into account the running yearly average\footnote{A 12 indices smoothing leads in IDL to consider 6 items before, the item itself, and the following 5 items giving rise to the rather strange looking asymmetry in the equation.}, i.e. summation and averaging of quantities from 6 months before, $t-6$, to 5 months after, $t+5$, the month of interest, where $t$ represents time as given by months):
\begin{equation}
\begin{aligned}
\frac{dN(t)}{dt}={}& a \sqrt{\frac{1}{12}\sum_{t-6}^{t+5}n_s(t-\delta t_1)} \\
&+\frac{b}{12} \sum_{t-6}^{t+5}{\frac{d \sum_{t-6}^{t+5}n_s(t-\delta t_2)/12}{dt}}\\
&+ c \left(\frac{1}{12}\sum_{t-6}^{t+5}N(t) - N_0\right).  
\label{compl}
\end{aligned}
\end{equation}
Due to the temporal averaging we cut off the first and the last months of our time series and fitted the period from May 2007 to August 2015. 
The obtained fitting parameters are listed in Tab. \ref{tab1}. For checking the stability of the model parameters we have redone the fitting also for a more limited data set where we only used the period from July 2008 until May 2014. While the model parameters do change quite considerably, the overall possible predictions due to the model are still acceptable and by far better than by a simple polynomial fit (see further below). The shifted input profiles for the square root of the relative sunspot number, $\sqrt{N_s}$, and the temporal derivative of the sunspot number are depicted in Fig. \ref{fig11}. The result of the modeling can be seen in Fig. \ref{fig12}. Here the detected change in number of MBPs (running yearly average) is shown with crosses and the obtained best fit by solid line.

\begin{table*}
	\centering
	\caption{gives the best fitting coefficients found for the general fitting of the dN/dt behaviour for the temporal change of the detected MBPs measured at the disc centre for a yearly running smoothing of all parameters.}
		\begin{tabular}{c|c|c|c|c|c|c}
		\hline
		\hline
		modeling period & a [month$^{-1}$] & b & c [month$^{-1}$] & $N_0$ & $\delta t_1$ [month] & $\delta t_2$ [month] \\
		\hline
		May 2007 / August 2015 & 2.728 & 0.962 & -0.018 & 255 & 22 & 8  \\
		July 2008 / May 2014& 2.575 & 1.516 & -0.038 & 448 & 23 & 10  \\
		\hline
			
		\end{tabular}
		
		\label{tab1}
\end{table*}

\section{Predictive capabilities of the model}

\begin{figure}
 \begin{center}
  \includegraphics[width=0.9\columnwidth]{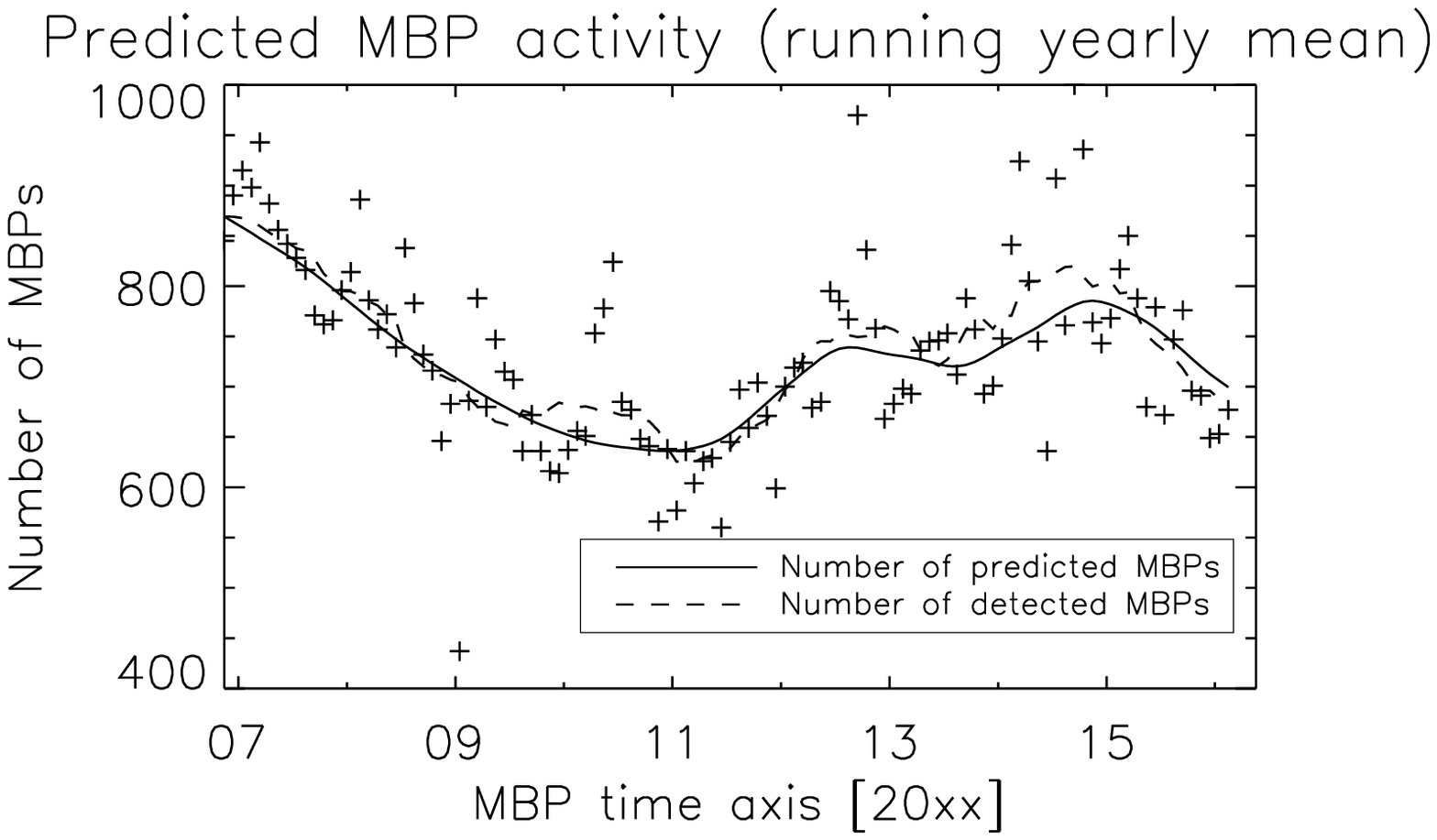}
 \end{center}
 \caption{the actual measured number of MBPs at disc centre (crosses) together with an annual running smooth of the actual data dashed line and the predicted number of MBPs at disc centre (solid line) obtained via a single starting number of MBPs and the fitting parameters listed in Table \ref{tab1} and Eq. \ref{compl}.}\label{fig13}
\end{figure}

After obtaining a well fitted $dN/dt$ curve we can use the obtained information to try to predict the variation of the number of MBPs at the disc centre by a simple recursive equation:
\begin{equation}
N(t)=N(t-1)+\Delta N(t-1),
\end{equation}
where $N(t)$ is the number of MBPs at disc centre and $\Delta N(t-1)$ can be calculated via Eq. \ref{compl} and is in principle only dependent on $N_s(t)$ the number of sunspots, its derivative, and the fitting parameters. Using this equation and a chosen starting value\footnote{We used November 2006, the first month of available Hinode data.} we calculated the predicted number of MBPs due to the given sunspot number as well as its change with time (running annual mean values). The result is shown in Fig. \ref{fig13}. Here the predicted number of MBPs is shown in solid line. For comparison reasons the actual number of MBPs (running annual mean) is shown in dashed line and the single real measurements are depicted as stars. The correlation between the predicted number of MBPs and the actual measured MBPs (running 12 month average) reaches a value of 0.959.

\begin{figure}
 \begin{center}
  \includegraphics[width=0.9\columnwidth]{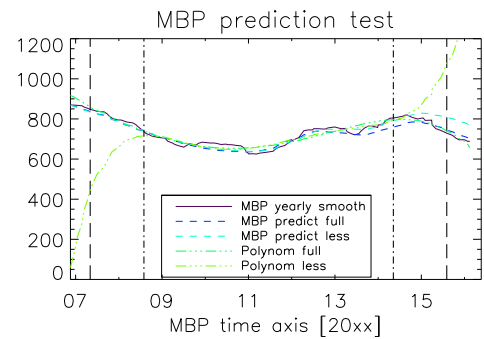}
 \end{center}
 \caption{the truly measured evolution of MBPs at disc centre after application of a yearly running average (solid line) together with two predictive models according to Eq. \ref{compl} and two simple 5th order polynomial fits. The vertical dashed lines give the used data range for the full fitting and the limited fitting.}\label{comparison}
\end{figure}

To prove the usefulness of our model we want to compare the predictive capability of our empirical model with a standard polynomial fit of 5 order (6 free fitting parameters). The result can be seen in Fig. \ref{comparison}. Here we show the running yearly average number of measured MBPs in full line, two different predictive models in blue dashed lines (for the dark blue model the maximum data range was used, for the cyan model the limited data range from July 2008 to May 2014), and two 5th order polynomial fits in green color in dashed dotted lines (again, one model created by the full data range, and the other one with the limited data set). While the polynomials in general correlate as good with the measured data as the predictive model (see also Tab. \ref{tab2}), when we come to predictions, they completely start to fail, while our empirical models can still follow quite closely the true evolution.

\begin{table*}
	\centering
	\caption{
	\begin{flushleft}
	gives the correlation coefficients between the measured evolution of MBPs, the predicted MBP evolutions, and via polynomial modeled MBP evolutions.
	\end{flushleft}
	}
		\begin{tabular}{c|c}
		\hline
		\hline
		correlation type & value  \\
		\hline
		predictive model full data range& 0.9586  \\
		predictive model small data range& 0.9427   \\
		predictive model small data range extended to full period& 0.9028   \\
		\hline
				polynomial model full data range& 0.9515  \\
		polynomial model small data range& 0.8986   \\
		polynomial model small data range extended to full period& -0.2588  \\
		\hline
			
		\end{tabular}
		
		\label{tab2}
\end{table*}

			
			
			
\section{Discussion}
In Sect. 3 we have seen that there is in general a good correlation between the number of sunspots and the number of MBPs detected at disc centre. However, this correlation only shows up when one shifts the two time series to each other. A question arising is, if it would be possible that this established correlation is just an artifact (e.g., due to instrumental aging)?

We were discussing this possibility at length already in our previous work \citet{2016A&A...585A..39U}. In that work we found that in the beginning a decrease of the instrument quality happens, which is largely offset by permanent refocusing attempts of the Hinode operation team\footnote{On a regular basis, about once a month, the Hinode team is measuring the optimum focus position and thus correcting for long term changes of the focal position due to ageing.}. Moreover, we saw that after several years the quality of the instrument starts to stabilise and that there should be no influence of the instrument anymore on the quality of the data. Our conclusion in this previous work, which we would like to keep upright, was that the overall behaviour of the measured evolution of the number of MBPs at disc centre is due to physical reasons and created by physical processes in the Sun. However, the exact numerical values can be slightly altered due to the changing instrument quality. Besides of that it is totally reasonable from the physical point of view that the number of sunspots will influence the activity and strength of the magnetic network. When a sunspot starts to decay, it will supplement and feed magnetic flux via the so-called moving magnetic features \citep[see, e.g.,][]{1973SoPh...28...61H} into the magnetic network. Thus it is highly plausible that there is indeed a temporal shifted correlation between the number of sunspots and the number of MBPs at disc centre (which represent strong magnetic flux elements of about 1 kG). Of course, at the end a correlation between two quantities can never proof that there is a true link between two quantities or a real process relating them. This could be only proven to a higher extend by direct and detailed investigations of the transport of flux elements, e.g., by investigating in full detail the flow maps in and around active regions \citep[e.g.,][]{jose}. However, it is a first strong indication for such processes. Moreover, from the viewpoint of physics, such a linkage between magnetic network activity and sunspot activity is very plausible and also used in theoretical solar cycle modeling \citep[e.g.,][]{2012ApJ...757..187T,2014ApJ...796...19T}. 

Coming back to the correlation between the sunspot number and the number of MBPs at disc centre, we would like to update our earlier work in the sense that we were actually now able to establish that there is not one single good correlation between the number of MBPs at the solar disc centre and the number of sunspots but that indeed the MBP cycle is broken in two parts. The decreasing phase of solar cycle 23 is shifted in time compared to the MBP decreasing phase by nearly 2 years (exactly 22 months) while the increase follows the rising phase of cycle 24 nearly co-temporal (as the correlation is already high for small time shifts but reaching a local maximum after 11 months). 

How can these temporal shifts plus the ones from the modeling now be explained or justified \citep[see also discussions in][]{2012ApJ...757..187T,2014ApJ...796...19T,2016A&A...585A..39U}?

We have to keep in mind that we observe the number of MBPs at the disc centre, which means very close to the equator\footnote{Due to the solar tilt angle to Earth´s ecliptic plane the actually observed solar latitude, by taking images at the apparent solar disc centre, varies slightly ($\pm 7^{\circ}$) during the year.}, while sunspots are emerging more or less in a latitude band between $\pm 30^{\circ}$ in the beginning of the cycle. Later on these zones of emergence mitigate closer and closer to the equator. Finally, at the end of the cycle, they can appear practically at the solar equator itself. Thus, the magnetic elements, created during the decay of the sunspots, need to be transported by distances as far reaching as $30^{\circ}$ from their emergence locations towards the equator. Only then they can be detected and show up visible in the analysed data set. This already explains, from a qualitative viewpoint, the time shifts seen in the rising phase as well as in the fitting parameters. For a more quantitative, but still basic, first principle analysis, we would now need to consider the available physical processes for such a magnetic flux transport. Among them are meridional flows, which would, however, rather transport the magnetic field away from the equator and to the poles with speeds of up to 10 m/s \citep[e.g.,][]{Hathaway}. Besides of such meridional flows we can also imagine that evolving supergranular cells distribute and diffuse magnetic fields. For our first principle analysis we would like to consider evolving supergranular cells as pushing magnetic elements on the solar surface in a simplified 1 dimensional\footnote{We are only interested in the latitudinal component.} random walk manner\footnote{That this is a reasonable approach can be seen by the fact that the behaviour of small-scale magnetic fields follow also on granular scales random walks \citep[e.g.,][]{utz}.}. Having a look on characteristic scales of supergranular cells we find the following values: sizes from 20 to 63 Mm with an average size of 36 Mm and turn over, or lifetimes, of about 1.7 days \citep[see, e.g., the review by][]{rieutord}.


Having now a look on the theory of a simple one dimensional random walk process, one can obtain the equation for the expectation value of travelled distance (in normalised characteristic distances) after a certain number of normalised (time) steps (approximated for large n) as $E(|n|)=\sqrt{(2 \cdot n /\pi)}$ \citep[see, e.g.,][]{wolfram}. In our case the normalisation of the stepsize is the typical lifetime and thus we can replace on the right side the step counting number n by $t_t/t_c$, where $t_c$ is the characteristic lifetime of the supergranulation and $t_t$ is the necessary transportation time. Similarly the necessary expectation value $E(|n|)$ is equivalent to the number of steps necessary to transport the magnetic fields over the necessary distance $d_t$ and thus given in multiples of the characteristic travel distance per time $d_c$. Replacing now these mentioned parts of the equation we end up with:
\begin{equation}
d_t/d_c=\sqrt{\frac{2\cdot t_t}{\pi \cdot t_c}},
\end{equation}  
or resolved for the necessary transportation time:
\begin{equation}
t_t=(d_t/d_c)^2\cdot \pi/2 \cdot t_c;
\end{equation}
But what are now the characteristic times and distances?

For the correlation between the decreasing phase of the sunspot cycle and the number of MBPs the delay happens most likely due to the fact that time is needed for the magnetic field to be transported out of the observable range. Thus, to estimate the time needed to clean the region close to the equator from magnetic elements originating from the sunspots, we can start by assuming that the last sunspots of the cycle were formed at the equator and that the magnetic field needs to be transported as far away as we could detect MBPs. Thus we can argue that these magnetic fields needs to be transported across the observable position furthest away from the equator, which would be the top boundary of the FOV at maximum solar tilt ($3.3^\circ+7.5^\circ\approx 11^\circ$). The second necessary parameter is now the charactristic distance travelled per turn-over time of a supergranule. For that we can assume that the elements are transported with a characteristic scale of around 9 Mm (half typical radius of the supergranulation\footnote{The factor half was chosen as not every supergranule will transport all elements always by a full radius but rather on average by about half of its radius.}). Inserting these values in the above equation we would end up with 19.8 months which fits quite well with the found 22 months of time lag. The detailed values for this calculation, its parameters, as well as for the other time shifts can be found in Table \ref{tab3}. 

For the rising phase, we assume in one case that the FOV is situated due to the solar tilt closest to the onset of the sunspot belt (depicted as rising phase max elongated FOV position case) while we also had a look on a more realistic set-up where we would assume that newly arrived elements ($30^\circ$ latitude) would be required to move inside the FOV of Hinode while the satellite is pointing at the equator. For the characteristic displacement we would have assumed the radius of the maximum size of the supergranules. In this way we make sure that these are really minimum time lags and that under normal conditions the true time lag should be rather larger than estimated here. In the first case we would end up with times of 5.1 months while the second one gives 10.8  months which agrees very well with the found 11 months time lag. Also we have seen in the first part of the paper that the rising phase is in addition highly correlated already when smaller time shifts are applied which can be due to the FOV of the satellite being more favorable positioned (as we have seen above). Moreover, the correlation coefficients are of course influenced in addition by the later cases of sunspot emergence, which will happen closer to the solar equator. 

To validate the model time shifts with this first principle analysis we assume that the flux needs to be transported from an average position ($15^\circ$) to the equator. The average position was chosen as the model should be valid for the whole solar cycle and thus the sunspots would be on average at a latitude position of the mentioned 15$^\circ$. For the $\delta t_2$ parameter we would assume fast movements as it relates to the immediate change in number of MBPs due to new emerging flux. Thus we would have assumed that the elements get pushed by a full average supergranular radius each time. The result would amount to 8.8 months which agrees again quite well with the obtained 8 months. The last parameter $\delta t_1$ measures the time lag to the sunspot number itself. As we have seen, and also know from theory, a movement by random walk is quite dispersive and thus we would have first elements reaching a certain position long before the bulk of features would come to a similar distance. Thus we estimated, for obtaining $\delta t_1$, the principle time when first elements could arrive, if they would be so luckily to be pushed all the time by the largest possible supergranules. This amounts to 3.2 months. Moreover, we estimated the time, it would take, for the end of the bulk of elements to arrive, by assuming that the elements would be always only pushed by half of the radius of a mean sized supergranular cell. That time would amount to 38.9 months. Thus, the first elements should arrive after 3.2 months while after 38.9 months the last elements should have arrived and therefore the influence from a certain sunspot should be over. Thus the average time lag between a sunspot arriving on the solar surface and its influence in our model should be the average of these two numbers amounting to ~21 months which is again in good agreement with the obtained 22 months in the model.

\begin{table*}
	\centering
	\caption{
	\begin{flushleft}
	gives expected time shifts on a first principle random walk analysis (for details see text).
	\end{flushleft}
	}
		\begin{tabular}{c|c|c|c}
		\hline
		\hline
		parameter & distance [$^\circ$]; [Mm] & characteristic displacement scale [Mm] & time [months] \\
		\hline
		rising phase max elongated FOV position& 20; 240 & 31.5 & 5.1 \\
		rising phase centre FOV position& 27.5; 350 & 31.5 & 10.8 \\
		\hline
		decreasing phase& 11; 135 & 9 & 19.8 \\
		\hline
		travel time fastest elements for mean distance& 15; 185 & 31.5 & 3.2 \\
		end of bulk travel time for mean distance& 15; 185 & 9 & 38,9 \\
		mean delay time for mean distances ($\delta t1$) & --- & --- & 21\\
		\hline
		tavel time for fast elements mean distances ($\delta t2$) & 15; 185 & 18 & 8.8 \\
		\hline

		\end{tabular}
		
		\label{tab3}
\end{table*}

Thus we can conclude from the observational correlation analysis, the modeling, and the interpretational random walk model findings that the injection of new flux coming from the sunspot cycle into the magnetic network happens on a faster temporal scale than the reduction of magnetic flux. Thus the diffusion mechanism which weakens and finally disperses these magnetic fields at the solar disc centre, before the magnetic flux ultimately gets transported to the poles, is acting slower than the injection mechanism.

In the next Sect. 4 we followed up with a more quantitative investigation into the formation and disintegration of MBPs and their correlation with the sunspot cycle. We introduced a model which can reproduce the temporal change of the number of MBPs. The fitting parameters of this model, as depicted in Tab. \ref{tab1}, are very interesting as they are related to physical processes and mechanisms. 

A plausibly interpretation of the $N_0$ parameter is that it represents a possible acting small-scale solar surface dynamo which permanently induces new magnetic fields and thus also some magnetic field enhancements independently of the sunspot activity. Hence there should be some permanent and constant number of MBPs visible, even when there is no global field acting, like during the extended minimum of cycle 23 \citep[e.g.,][]{2011SoPh..274...87M}. These constantly created background MBPs would be exactly described by $N_0$. 
Similarly, the parameter $c$ could be seen as related to the large-scale flow pattern (meridional flows) as it governs the speed of reduction of magnetic flux (number of MBPs). A possible enhanced flow field could destroy flux concentrations more rapidly and in addition move magnetic flux out of the field of view faster (disperse the field in the direction of the poles). Then there are the parameters $a$ and $b$ which describe together with $\delta t_1$ and $\delta t_2$ the coupling of the sunspots to the magnetic network as seen by MBPs. There are two striking features with these parameters. First of all we recover the temporal shifts from the previous data analysis with 22 months for the sunspot coupling (the temporal shift in the decreasing phase of cycle 23) and 8 months (which is close to the one year for the increasing phase of cycle 24) for the monthly derivative of the sunspot number. As the rate of change in the sunspot number is related directly to fresh magnetic flux injection we get another indication that the flux injection from sunspots works much faster than the removal of magnetic flux from the network. The other striking feature is that the coupling parameter $a$ and the temporal shifts are very stable parameters in regards of changing the data range for creating the predictive model, while the other parameters $b$, $c$, and $N_0$ seem to vary in a higher degree and are less well determined. This has also to do with the kind of fitting as e.g., $c$ and $N_0$ are counteracting parameters. A larger background flux density $N_0$ could be compensated by a larger meridional flow and magnetic field dispersion related to $c$. Thus, as these parameters couple, they can be not estimated that easily and with high accuracy.  

At this point we would like to make clear that the developed model creates a good representative and prediction capability for the observed roughly 10 year period, however, it is a) just a toy model based on phenomenological terms \citep[for an older, in some way similar approach, see, e.g.,][]{1994SoPh..150....1S}, and b) the parameters have to be taken extremely carefully as they represent only a particular (nevertheless useful) local minimum solution to the multitude of possible fitting solutions for a 6 parameter model. In the future it would be necessary to compare the findings with other, more direct methods, like flow field analysis \citep[e.g., the local correlation technique, see][]{1988ApJ...333..427N}, which could also yield insights in the magnetic flux transport by plasma motions by comparing plasma flows with changes of the magnetic field and its position \citep[see, e.g.,][]{2008ApJ...679..900V}.

A main reason in the variation of the parameters $b,c,N_0$ with the covered period (see Tab. \ref{tab2}) will be due to the non separation of the hemispheric sunspot numbers in this study. When the Hinode satellite is observing the disc centre within its campaign, it is covering for half a year to a larger extend the equator and the region slightly northern of it (so the northern hemispheric sunspots will have a larger influence on the observed network) and for the other half of a year the equator plus the region slightly southern of it has greater influence \citep[so the influence will be to a higher degree coming from the southern hemisphere; see also ][]{2016A&A...585A..39U}. This is due to the fact that the apparent solar tilt angle varies between roughly $\pm 7.5^{\circ}$ while the total FOV of Hinode in latitude is about $6.6^{\circ}$. Thus for a more accurate model these effects should be considered in future models and might help to determine the parameters more uniquely as neglecting this effect will lead to a small systematic error (likely increasing the observed time lags) and more likely to larger erros in the estimated time lags. The theoretical interpretation of the parameters is outstanding but should be quite interesting as the decay constant should be coupled to the meridional flows dispersing and transporting the magnetic field from the centre to the poles and the background number of MBPs will be related to an acting surface dynamo and thus could tell us in principle about the strength of it. Thus a more thorough analysis in the future could yield interesting new and additional insight.

\section{Conclusion}
In this work we had a detailed look on the evolution of MBPs close to the solar disc centre by investigating the Hinode synoptic data set which was recorded over a period of roughly 10 years and is the most suitable data set available in the world for such a purpose as it is stable and seeing free due to the space conditions. The evolution of the number of MBPs at the disc centre can tell us about the important coupling mechanism between the small-scale magnetic fields and the large scale solar cycle induced sunspot cycle as well as about magnetoconvection processes, especially the magnetic flux transport.
We have shown that there is not a simple unique coupling between the variation of MBPs but that actually the temporal shift between the minimum in the sunspot cycle and the minimum in the number of MBPs is different from the one obtained when one investigates the shift between the corresponding maximum values. This means that the injection of new magnetic flux from the global field dynamo works on a faster time scale than the reduction of the magnetic flux from the magnetic network, which is probably due to plasma flows dispersing the magnetic field and transporting it away from the disc centre. In the second part of this work we developed a coupling model between the temporal change of the number of MBPs at the disc centre and the sunspot number. Detailed investigations of this model have shown that characteristics like the decay constant (in this work denoted as $c$) and the background magnetic field constant ($N_0$) cannot easily and with full stability obtained by the enormous but yet limited data set. However, the predictive capability of the introduced empirical model still gives correlations better than 0.9 for periods extended by roughly 2 years anterior and posterior to the used data set. Thus, future work on such a model seems promising and probably a huge step forward can be achieved by separating the hemispheric sunspot numbers. In addition we should keep in mind that ephemeral regions are also a main source for the injection of magnetic flux into the network, wherefore it would be good to include them into the modeling efforts. Finally, we would like to remark that the observed temporal lags between the sunspot cycle and the detected MBPs close to the solar equator might be explained by a simple random walk model were supergranules push and move the magnetic elements. 


\begin{ack}
This research received support by the Austrian Science Fund (FWF): P27800. Additional funding was possible through an Odysseus grant of
the FWO Vlaanderen, the IAP P7/08 CHARM (Belspo), and GOA-2015-014
(KU Leuven). Besides, the authors are very thankful to the Hinode team for operating and maintaining the extremely fruitful Hinode space mission over all these years. Especially the lead author D.U. wishes to express his gratitude to the whole Hinode team as this mission was his starting point and entry point into the solar physics community more or less exactly 10 years ago.
Hinode is a Japanese mission developed and launched by ISAS/JAXA, collaborating with NAOJ as a domestic partner, NASA and STFC (UK) as international partners. Scientific operation of the Hinode mission is conducted by the Hinode science team organized at ISAS/JAXA. This team mainly consists of scientists from institutes in the partner countries. Support for the post-launch operation is provided by JAXA and NAOJ (Japan), STFC (U.K.), NASA (U.S.A.), ESA, and NSC (Norway). We wish to express our gratitude to the anonymous referee, who, due to his criticism of the model, enabled us to strengthen our modeling approach considerably.
\end{ack}

\end{document}